\documentclass[aps,prl,reprint,groupedaddress,showpacs]{revtex4-1}
\usepackage{hyperref}

\usepackage{graphicx}
\usepackage{epsfig}
\usepackage{amsmath}

\begin{document}

\title{Ramsauer effect in one-dimensional quantum walk with multiple defects}
\author{Ho Tat Lam}
\author{Kwok Yip Szeto}
\email[Corresponding Author: ]{phszeto@ust.hk}
\affiliation{Department of Physics, Hong Kong University of Science and Technology, Hong Kong}

\begin{abstract}
Experimental observations of quantum walks in one dimension have provided many exciting applications in quantum computing, while recent theoretical investigation of single phase defect in these system points towards interesting phenomena associated with bounds states. Here we obtain analytical solutions of quantum walk with a general quantum coin in one dimension with multiple defects, with new prediction on the condition for zero reflectance for scattering state, and the existence of an analogy to the Ramsauer effect for multiple defects. We also show the transition from the zero reflectance state to the bound state can provide a method for preparing the quantum walk in a bound state. Applications to systems similar to thin film optics are suggested.   
\end{abstract}
\maketitle

After Feynman's proposal of quantum cellular automaton \cite{1,2}, researchers have introduced a model of discrete time quantum walk\cite{3,4,5} which has since become a subject of intense research interest. Theoretically it is a model for universal quantum computation\cite{6,7} and quantum algorithm\cite{8,9,10,11}, with relevance in many subjects, such as the topological phases\cite{12}, quantum phase transition and localization in optical lattices\cite{13,14}, energy transfer in photosynthetic systems\cite{15,16}, and breakdown of electric field driven systems\cite{17}. Experimental realizations of the discrete time quantum walk are observed in optical resonator\cite{18}, nuclear magnetic resonance\cite{19}, trapped ions\cite{20,21,22} and neutral atoms\cite{23}, photons in different structures such as beam splitter array, fiber loop and waveguide\cite{24,25,26,27,28,29,30,31}.

In this paper we find an analog of the Ramsauer effect in one-dimensional quantum walk with multiple defects. The Ramsauer effect in quantum mechanics refers to the phenomenon of the probability of electrons tunneling through potential barrier oscillates with respect to energies, and the reflectance vanishes at several different energies due to the interference between the incoming wave and the reflecting waves. For one-dimensional quantum walk with defects, we solve for the scattering states as well as bound states and find similar oscillatory behavior in the reflectance.

The simplest soluble model of discrete-time quantum walk\cite{3,4,32,33,34,35,36,37,38} is described by iteration of a unitary transformation acting on a two-state particle. The particle is represented by a wave function $|\psi\rangle=\sum(a_n|L\rangle+b_n|R\rangle)|n\rangle$ where $|n\rangle$ is the state that quantum walk localized at position $n$ whereas left $|L\rangle$ and right $|R\rangle$ states represent the two internal degree of freedom. Here, $a_n$ and $b_n$ are the left and right components of the wave function at position $n$. The unitary transformation $U = S{\bullet}C$ consists of a quantum coin operation $C$ that acts on the internal state, followed by a shift operator $S$ to evolve the particle coherently in position space. The shift operator $S$ is $S = |L\rangle{\langle}L|\sum|n\rangle{\langle}n+1|+|R\rangle{\langle}R|\sum|n\rangle{\langle}n-1|$ which moves the left and right states to the left and right direction by 1 unit respectively. A common example of the coin operator $C$ is Hadamard coin given by,
\begin{equation}
C= \frac{1}{\sqrt{2}}\left( \begin{array}{ccc}
1 & 1 \\
1 & -1 \end{array}\right)
\end{equation}
where the matrix is written in the basis of $\{|L\rangle,|R\rangle\}$. A general coin operator is a rotation matrix parametrized by a coin parameter $\theta$ as,
\begin{equation}
C=\cos\theta\sigma_z+\sin\theta\sigma_x=\left( \begin{array}{ccc}
\cos\theta & \sin{\theta} \\
\sin{\theta} & -\cos{\theta} \end{array} \right)
\end{equation}
When $\theta = \pi/4$, the general quantum walk reduces to the Hadamard Walk. The simplest solution for one-dimensional quantum walk is plane waves,
\begin{equation}
\psi(n,t) = e^{-iEt+ikn}\left(\begin{array}{ccc}
a_k\\
b_k \end{array} \right)
\end{equation}
where $E$ and $k$ are the quasi-energy and quasi-momentum with period of $2\pi$ satisfying,
\begin{equation}
{\sin}E = -{\cos}\theta{\sin}k\label{Eq. 7}
\end{equation}
and $a_k$ and $b_k$ are the left and right components of the wave function given by,
\begin{equation}
\left(\begin{array}{ccc}
a_k\\
b_k \end{array} \right)
= \frac{1}{\sqrt{2-2\cos\theta\cos(E-k)}}\left(\begin{array}{ccc}
e^{ik}\sin\theta\\
e^{-iE}-\cos{\theta}e^{ik} \end{array} \right)
\end{equation}
For each $E$, there are two possible quasi-momentum $k=-\arcsin(\sin E/\cos\theta)$  and $k=\pi+\arcsin(\sin E/\cos\theta)$ representing waves going to the direction of increasing $n$ or decreasing $n$.
The dispersion relation in Eq.~\ref{Eq. 7} is unusual in the sense that the quasi-energies are antisymmetric with respect to quasi-momentum. This antisymmetry is due to the fact that the coin can never recover the identity matrix by varying $\theta$ parameter, instead, it is reduced to $\sigma_z$ matrix when $\theta=0$. When $\theta=0$, the left and right component decoupled. The left component moves to the left without any changing after one unitary transformation which is equivalent to a wave with quasi-momentum $k=0$ and quasi-energy $E=0$ while the right component moves to the right and consistently changes its sign each step which is equivalent to a wave with quasi-momentum $k=\pi$ and quasi-energy $E=0$. Thus, the wave with $k=0$ has a negative group velocity and the wave with $k=\pi$ has a positive group velocity when the quasi-energy $E=0$. For $\theta\in(-\pi/2,\pi/2)$, $\cos\theta$ is positive, thus the group velocity$v_g=dE/{dk}$=$-\cos{\theta}\cos k/\cos E$ is positive for $k\in(\pi/2,3\pi/2)$ and is negative for $k\in(-\pi/2,\pi/2)$ if we restrict all the discussion to $E\in(-\pi/2,\pi/2)$. On the other hand, for $E\in(\pi/2,3\pi/2)$, the group velocity changes its direction. In the later discussion, we only study the branch of $E\in(-\pi/2,\pi/2)$.

An extension of this one-dimensional quantum walk model to include a single defect that modifies the phase of the state at the origin $|0\rangle$ was recently investigated by Wojcik \textit{et al}.\cite{39}. Many papers have since been written on this subject,\cite{40,41}. Each time the walker passes through the phase defect $\omega=e^{i\phi}[\phi\in(0,2\pi)]$ at the origin, it will gain an extra phase $e^{i\phi}$. The unitary transformation with phase defect is $U_{\phi}=S{\bullet}e^{i\phi\delta_{0,n}}C$ and the evolution of this state is governed by $|\Psi(t)\rangle=U_\phi^t|\Psi(0)\rangle$.

We first study the scattering state when the incoming wave comes from the negative side, i.e. $k\in(\pi/2,3\pi/2)$. The defect is treated as a boundary between scattering waves, which is similar to the situation in thin film optics\cite{42}. After hitting on the phase defect, part of the wave transmits through the defect with transmitting amplitude $t$ and the other is reflected with reflecting amplitude $r$ calculated in Appendix A.
\begin{widetext}
\begin{equation}
r=\textsl{sign}(k-\pi)\frac{(1-e^{i\phi})\sqrt{\sin(k-E)\sin(k+E)}(e^{i\phi}-e^{-2iE})}{e^{i\phi}[(e^{i\phi}-1)\sin(E-k)+i(e^{-2iE}-\cos2k)\cos\theta]+\sin(E+k)(e^{i\phi}-e^{-2iE})}\label{Eq. 10}
\end{equation}
\begin{equation}
t=\frac{-\sin(2k)\cos^2{\theta}e^{i(\phi-k-E)}}{e^{i\phi}[(e^{i\phi}-1)\sin(E-k)+i(e^{-2iE}-\cos2k)\cos\theta]+\sin(E+k)(e^{i\phi}-e^{-2iE})}\label{Eq. 11}
\end{equation}
\end{widetext}
where $E$ is a given by Eq.~\ref{Eq. 7} and the sign function is $+1$ for positive variable and $-1$ for negative variable. This sign function is responsible for selecting one of the two branches of the square root function. The unitary transformation $U_\phi$ is invariant under the transformation $\theta\to\pi-\theta$, the exchange of basis $|L\rangle\leftrightarrow|R\rangle$ and space reversal transformation. The three transformations brings the first equation of the following evolution equations to the second and vice versa,
\begin{equation}
\begin{aligned}
a_{n-1}(t+1)=\cos\theta a_{n}(t)+\sin\theta b_n(t)\\
b_{n+1}(t+1)=\sin\theta a_{n}(t)-\cos\theta b_n(t)
\end{aligned}
\end{equation}
Thus the reflecting amplitude, transmitting amplitude and the wave function for the left going wave can be obtained from the right going wave by change $\theta$ to $\pi-\theta$.

\begin{figure}
\includegraphics[scale = 0.4]{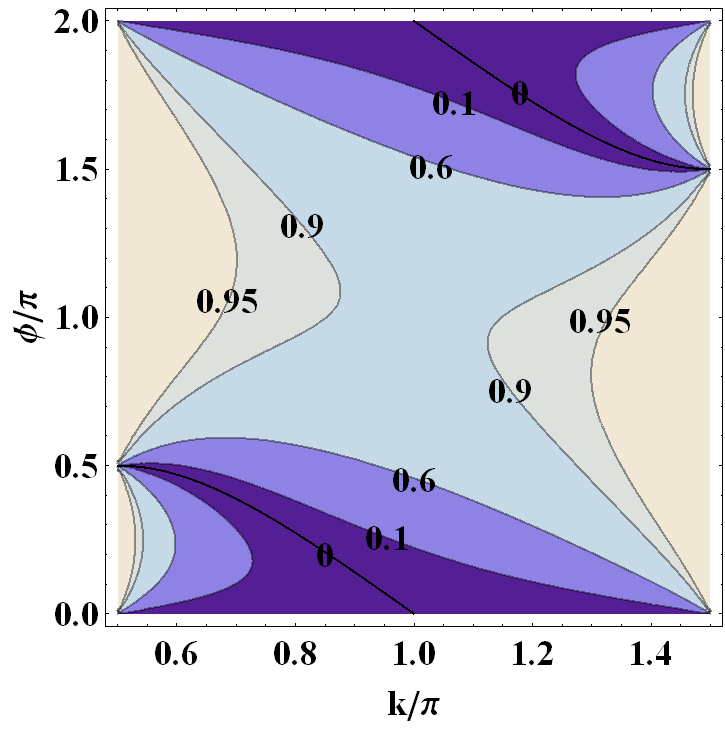}
\caption{\label{Figure 1}Contour plot of the reflectance of Hadamard walk versus phase shift $\phi$ and quasi-momentum $k$. Darker color represents lower value of reflectance. }
\end{figure}

Note that $|r|^2+|t|^2=1$. In Fig.~\ref{Figure 1}, we illustrate the contour plot of the reflectance $|r|^2$ of a Hadamard walk, i.e. $\theta = \pi/4$, on the plane of the phase defect $\phi$ and quasi-momentum $k$.
When $\phi=0$, the quantum walk reduces to the situation with no defect and the reflecting amplitude $r$ vanishes. Interestingly, when $\phi=-2E$ corresponding to the 0 level curve in Fig.~\ref{Figure 1}, the reflecting amplitude also vanishes while the transmitting amplitude is $t=e^{-2ik}$. In this situation, quantum walk will transmit with zero reflectance and gain an extra phase $e^{-2ik}$ as if it passes through node $1$ coincides with node $-1$. Although the condition for zero reflectance, $\phi=-2E$, is independent of the coin parameter $\theta$, not all phase defects produce zero reflectance phenomena for a given $\theta$. The zero reflectance occurs only when the defect phase is $\phi=-2E=2\arcsin(\cos{\theta}\sin{k})$
and $k\in(\pi/2,3\pi/2)$, thus only for $\phi\in(-\pi+2\theta,\pi-2\theta)$. For $\phi\in(\pi-2\theta,3\pi-2\theta)$, the reflectance is nearly 1. This range of $\phi$ for zero reflectance agrees with the 0 level curve in Fig.~\ref{Figure 1} where zero reflectance occurs only in $(-\pi/2,\pi/2)$. We define a critical value $\phi_{\textsl{critical}}=\pi-2\theta$ for the defect phase $\phi$ at a given $\theta$, so that zero reflectance can occur when $-\phi_{\textsl{critical}}<\phi<\phi_{\textsl{critical}}$. One can therefore select a particular scattering state that has zero reflectance by tuning the phase of the defect, making the defect as a state selector.

\begin{figure}
\includegraphics[scale = 0.555]{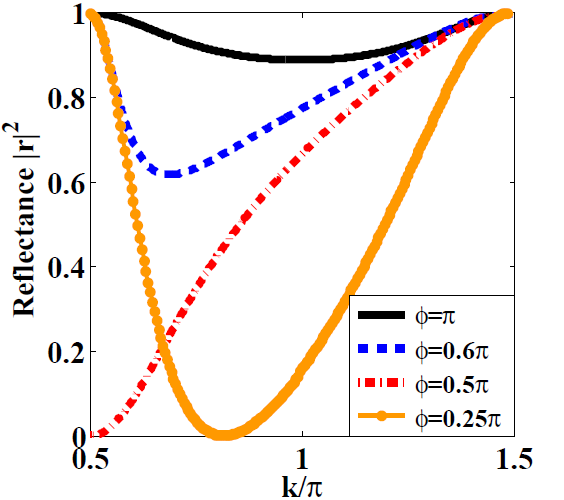}
\caption{\label{Figure 2}Reflectance of Hadamard walk versus the quasi-momentum $k$ for different $\phi$.}
\end{figure}

\begin{figure}
\includegraphics[scale = 0.55]{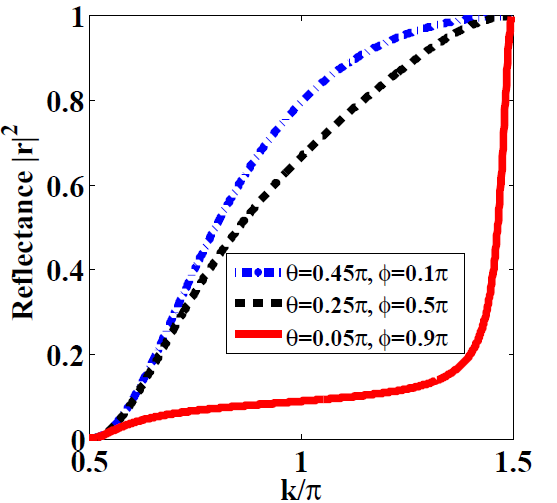}
\caption{\label{Figure 3}Reflectance for quantum walk with different $\theta$ at critical phase.}
\end{figure}

The quasi-momentum dependence of the reflectance $|r|^2$ for different $\phi$ of Hadamard walk is illustrated in Fig.~\ref{Figure 2}. For $\phi$ greater than $\pi/2$, the critical value for Hadamard walk, zero reflectance state disappears as predicted. Nevertheless, there is still a state with minimum reflectance, which magnitude increases with $\phi$. For $-\phi_{\textsl{critical}}<\phi<\phi_{\textsl{critical}}$, zero reflectance can be observed and as $\phi$ decreases, the zero reflectance quasi-momentum approaches 0 and the reflectance decreases in general. 
For non-Hadamard walk, there is a general dependence of the critical phase $\phi_{\textsl{critical}}$ on the coin parameter $\theta$. In Fig.~\ref{Figure 3}, we show the reflectance versus quasi-momentum at this critical phase $\phi_{\textsl{critical}}$ at different $\theta$. When $\theta=0.05\pi$, the critical phase $\phi_{\textsl{critical}}=0.9\pi$ and the overall reflectance is relatively small compared with Hadamard walk; while the critical phase for $\theta=0.45\pi$ is $\phi_{\textsl{critical}}=0.1\pi$ and the reflectance is relatively greater. The result shows that the overall reflectance at $\phi_{\textsl{critical}}$ increases when $\theta$ increases.

It has been shown that bound states may exist for certain $\phi$ in Hadamard walk\cite{39}. We now investigate the condition for the existence of bound state using general coin and its transition at critical phase $\phi$. For bound state, the quasi-momentum $k$ is complex so that the wave function decays with distance. Supposing $k=-i\kappa$, similar to the calculation for scattering state, by changing the $k$ into $-i\kappa$ and solving the equations we get four solutions. The details is presented in the Appendix B.
\begin{equation}
e^{\kappa_{m\pm}}=\frac{\pm i}{\sqrt{1+2\sin{\phi}\tan{\theta}[(-1)^m+\tan{(\phi/2)}\tan\theta]}}\label{Eq. 14}
\end{equation}
where $m=1,2$. Notice that we have obtained two pairs of bound states with same quasi-energy, $\kappa_{1\pm}$ and $\kappa_{2\pm}$ while Wojcik \textit{et al} only obtained two bound states\cite{39}. It is because they study the eigenstate of double-step evolution operator $U_{\phi}^2$ instead of single-step evolution operator $U_{\phi}$. For double-step operator, only the sites differed by distance of 2 are related. Then, $e^{\kappa_{1+}}$ and $e^{\kappa_{1-}}$ reduces to same bound state of $U_{\phi}^2$ because the decay rate after distance of 2 for both states are the same, i.e. $e^{2\kappa_{1+}}=e^{2\kappa_{1-}}$. So the four bound states of $U_{\phi}$ reduce to the two bound states of $U_{\phi}^2$. However, $\kappa_{1+}$ bound state and $\kappa_{1-}$ bound state are different at the odd number sites so they are different states. We therefore conclude that there exist four bound states in quantum walk with single phase defect. Note that bound state exists only when the decay constant $e^{\kappa}$ is less than 1, otherwise, the wave function diverges at infinity. From Eq.~\ref{Eq. 14}, it can be shown that $\kappa_1\pm$ bound state exists only when $2\pi>\phi> \phi_{\textsl{critical}} =\pi-2\theta$ and $\kappa_2\pm$ bound state exists only when $-2\pi<\phi<-\phi_{critical}=-\pi+2\theta$.

With both the scattering state and the bound state found, we can consider the transition from one to the other by tuning the defect phase adiabatically. Suppose the quantum walk is prepared in the $\kappa_{1+}$ bound state for $\phi > \phi_{\textsl{critical}}=\pi-2\theta$ and $\phi$ is tuned adiabatically across the critical value $\phi_{\textsl{critical}}$. The bound state will no longer exist but transforms into a right going wave with $k=\pi/2$ when $\phi<\phi_{\textsl{critical}}$. As $\phi$ decreases further, the state becomes the zero reflectance state. This observation is consistent with the fact that zero reflectance occurs only for $\phi\in(-\pi+2\theta,\pi-2\theta)$. Similarly, the ${\kappa}_{1-}$ bound state transforms into the left going wave with zero reflectance however with quasi-energy in $(\pi/2,3\pi/2)$. At $\phi=-\phi_{\textsl{critical}}=-\pi+2\theta$, the ${\kappa}_{2\pm}$ bound states transform into right and left going zero reflectance waves. It is possible that one can also obtain the bound state from the zero reflectance state by tuning $\phi$. This observation of the transition can be also seen from the similarity between the equation of zero reflectance state, Eq.~\ref{Eq. 25}, with $r=0$ and the bound state equation, Eq.~\ref{Eq. 30}. 

\begin{figure}
\includegraphics[scale = 0.2]{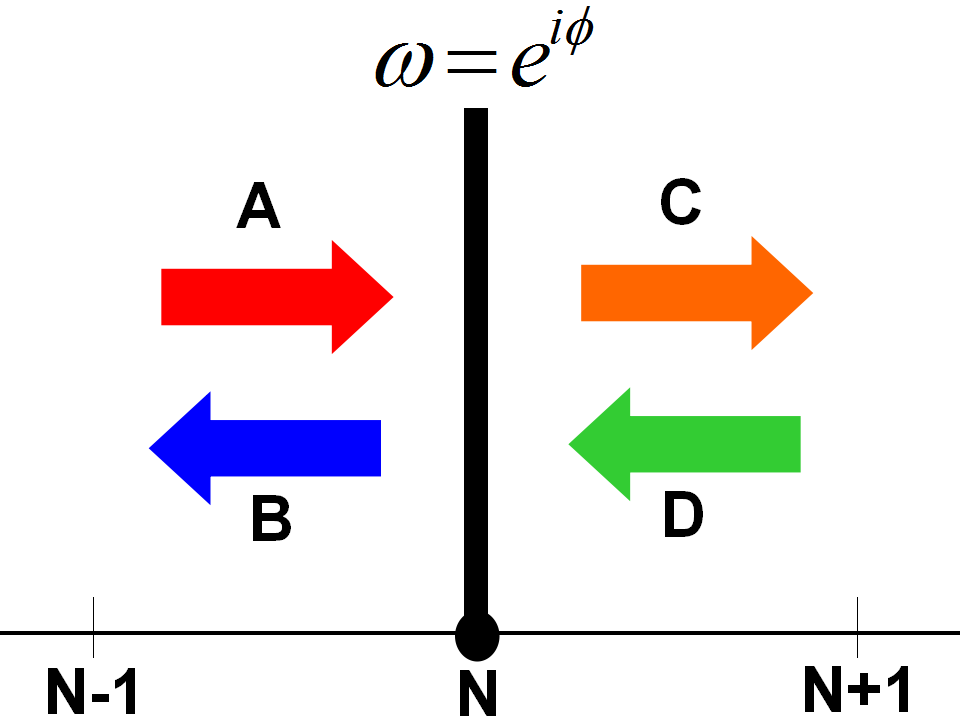}
\caption{\label{Figure 4}The amplitude of the wave function around a phase defect with phase shift $\phi$ at position $N$.}
\end{figure}

We now study the scattering state of multiple phase defects by first deriving the transition matrix and then apply it to the case with double phase defects. In multiple defects, the wave function is separated into several parts. Between two defects, the wave function is the superposition of left and right going wave with amplitudes $A$, $B$, $C$ and $D$ shown in Fig.~\ref{Figure 4}. 
When there is a phase defect at position $n=N$, the wave function between the defect and its neighbouring defects is,
\begin{equation}
\begin{aligned}
\psi(n) = Ae^{ikn}\left(\begin{array}{ccc}
a_k\\
b_k \end{array} \right)+Be^{i(\pi-k)n}\left(\begin{array}{ccc}
a_{\pi-k}\\
b_{\pi-k} \end{array} \right)&,(n<N)\\
\psi(n) = Ce^{ikn}\left(\begin{array}{ccc}
a_k\\
b_k \end{array} \right)+De^{i(\pi-k)n}\left(\begin{array}{ccc}
a_{\pi-k}\\
b_{\pi-k} \end{array} \right)&, (n>N)\\
\psi(N) = \left(\begin{array}{ccc}
Ce^{ikN}a_k+De^{i(\pi-k)N}a_{\pi-k}\\
Ae^{ikN}b_k+Be^{i(\pi-k)N}b_{\pi-k} \end{array} \right)&
\end{aligned}
\end{equation}
By matching the evolution equations at $n=\pm1$ and rearranging them, the amplitude $A$, $B$, $C$ and $D$ are connected by a matrix $\Lambda(\omega,N)$,
\begin{equation}
\left( \begin{array}{ccc}
C \\
D \end{array} \right)
=
\Lambda(\omega,N)\left( \begin{array}{ccc}
A \\
B \end{array} \right)
\end{equation}
\begin{widetext}
\begin{equation}
\begin{aligned}
\Lambda(\omega,N)= &\left( \begin{array}{ccc}
e^{-ikN}  & 0 \\
0 & e^{-i(\pi-k)N}\end{array} \right)
\left( \begin{array}{ccc}
{\omega}a_k & {\omega}a_{\pi-k} \\
(\omega-1){\tan\theta}a_k-b_k & (\omega-1){\tan\theta}a_{\pi-k}-b_{\pi-k} \end{array} \right)^{-1}\\
&\left( \begin{array}{ccc}
a_k-(\omega-1){\tan\theta}b_k & a_{\pi-k}-(\omega-1){\tan\theta}b_{\pi-k} \\
{\omega}b_k & {\omega}b_{\pi-k} \end{array} \right)
\left( \begin{array}{ccc}
e^{ikN}  & 0 \\
0 & e^{i(\pi-k)N}\end{array} \right)
\end{aligned}
\end{equation}
\end{widetext}
In a system with $n$ defects at $N_i$, $i=1,..,n$, after transmitting through all the defects, the transmitting amplitude $t$  and reflecting amplitude $r$ are connected through a series multiplication of $\Lambda(\omega_i,N_i)$.
\begin{equation}
\left( \begin{array}{ccc}
t \\
0 \end{array} \right)
=
\Pi^{n}_{i=1}\Lambda(\omega_i,N_i)
\left( \begin{array}{ccc}
1 \\
r \end{array} \right)
\end{equation}
By solving this system of linear equations, the transmitting and reflecting amplitude can be obtained. 

We now illustrate this with an example with two defects of same phase $\phi$ at position $n=0$ and $n=d$ respectively and an oscillatory reflectance that can be shown as an analogue to the Ramsauer effect. 
\begin{figure}
\includegraphics[scale = 0.38]{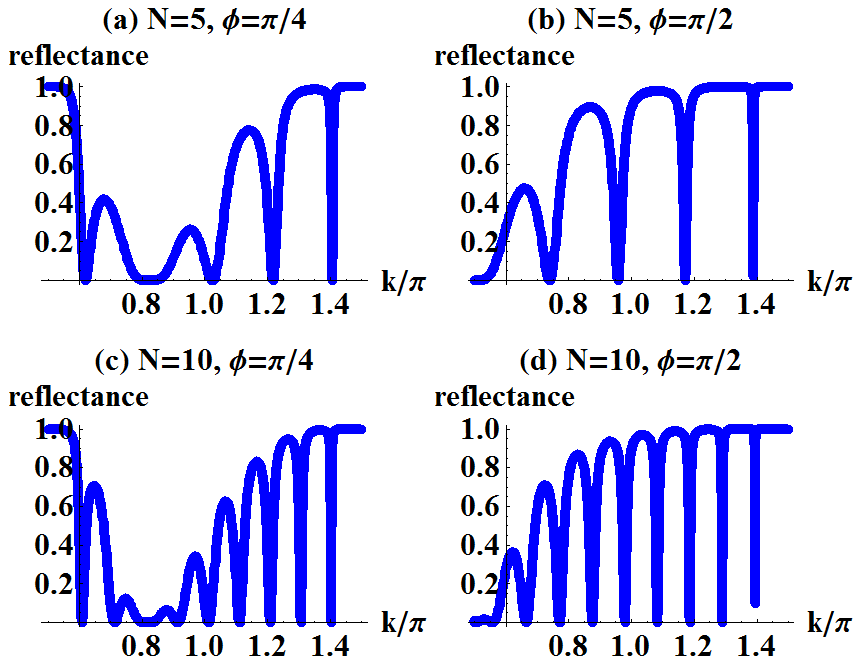}
\caption{\label{Figure 5}Reflectance of Hadamard walk going through two phase defects with same phase shift $\phi$ separated by distance $N$ versus the quasi-momentum $k$.}
\end{figure}
Fig.~\ref{Figure 5} shows the quasi-momentum dependence of the reflectance for Hadamard walk. There are three interesting phenomena. Firstly, the reflectance vanishes at more than one quasi-momentum similar to the Ramsauer effect in quantum mechanics, which says that electrons can perfectly tunnel through the potential barrier for several different energies due to the interference between the incoming wave and the reflecting waves. Secondly, the number of dips of reflectance versus quasi-momentum graphs generally increases with $N$. Thirdly, for $\phi$ from 0 to $\pi$, the larger the $\phi$, the higher the contrast in reflectance.

The similarity between our result for the reflectance of general quantum walk with multiple phase defects and Ramsauer effect suggests that we can actually treat the phase defects in quantum walk as an analogue of delta potentials. The evolution operator of the quantum walk with phase defect differs from evolution operator of the free quantum walk by a factor of $e^{i{\phi}\delta_{0,n}}$. It resembles the evolution operator with non-zero potential in quantum mechanics, $U(t)=e^{-iHt}=e^{-iV(n)t}U_{\textsl{free}}(t)$. This points to the interpretation that a phase defect localized at position $n$ is equivalent to a delta potential, $V=-\phi\delta_{0,n}$. This interpretation of defect explains the Ramsauer effect in the quantum walk system because it also exists in double delta potential.

One dimensional quantum walk with single phase defect has been solved analytically for the scattering as well as bound states, showing existence of zero reflectance, or total transmittance, when the phase of the phase defect lies in certain range of values. The transition between scattering state and bound state may be achieved through tuning the phase of the defect. A generalization of these results to multiple defects indicates that the reflectance has multiple minima at specific values of the quasi-momentum $k$, thus showing that preparation of the initial state of the quantum walker at several quasi-energy will yield zero reflectance. This provides a tool to select states of specific quasi-energy by suitably tuning the phase of the defects. With the recent formulation of quantum walk using two-component Dirac like Hamiltonian\cite{43}, our studies on phase defects can be extended to two and three dimensional lattice, decorated with periodic lattice of defects. Many phenomena in conventional solid state physics can be transposed into a problem of quantum walk by creating phase defects at different positions.
\begin{acknowledgments}
This work has been supported partially by  grant FSGRF13SC25 and FSGRF14SC28.
\end{acknowledgments}

\appendix*
\section{\label{Appendix A}Appendix A. Reflectance and transmittance of a scattering state for one defect}
The wave function of a scattering state is,
\begin{equation}
\begin{aligned}
\psi(n) &= e^{ikn}\left(\begin{array}{ccc}
a_k\\
b_k \end{array} \right)+re^{i(\pi-k)n}\left(\begin{array}{ccc}
a_{\pi-k}\\
b_{\pi-k} \end{array} \right) \quad (n<0)\\
&= te^{ikn}\left(\begin{array}{ccc}
a_k\\
b_k \end{array} \right)  \quad (n>0)
\end{aligned}
\end{equation}
At $n=0$, the left component $a_0$ follows the wave function at the positive region because it comes from the shifting of the wave function at $n>0$. Similarly, the right component $b_0$ follows the wave function at the negative region. Thus the wave function at $n=0$ is,
\begin{equation}
\psi(0) = \left(\begin{array}{ccc}
ta_k\\
b_k+rb_{\pi-k} \end{array} \right)
\end{equation}
The transmitting amplitude $t$ and the reflecting amplitude $r$ satisfies following system of linear equations,
\begin{equation}
\begin{aligned}
\omega[{\cos\theta}ta_k+{\sin\theta}(b_k+rb_{\pi-k})]
=&{\cos\theta}(a_k+ra_{\pi-k})\\&+{\sin\theta}(b_k+rb_{\pi-k})\\
\omega({\sin\theta}ta_k-{\cos\theta}(b_k+rb_{\pi-k}))=&t({\sin\theta}a_k+{\cos\theta}b_k)\label{Eq. 25}
\end{aligned}
\end{equation}
The first and second equations come from matching the left component at $n=-1$ and the right component at $n=1$ after one evolution with or without a phase defect at $n=0$ respectively. Solving these two equations, we obtain the reflecting amplitude $r$ and transmitting amplitude $t$ in Eq.~\ref{Eq. 10} and Eq.~\ref{Eq. 11}.
\section{Appendix B. Bound state for one defect}
The wave function of the bound states satisfies the boundary condition at the infinity $\lim_{n\rightarrow\infty}\psi=0$. Accordingly the quasi-momenta $k$ becomes a complex variable, thus we introduce a new variable, decay rate $\kappa=ik$. Then wave function of the bound state is as follows
\begin{equation}
\begin{aligned}
\psi(n) &= e^{{\kappa}n}\left(\begin{array}{ccc}
a_{\kappa}\\
b_{\kappa} \end{array} \right), (n<0);\quad \left(\begin{array}{ccc}
ta_{i\pi-\kappa}\\
b_{\kappa} \end{array} \right), (n=0)\\
&= te^{(i\pi-\kappa)n}\left(\begin{array}{ccc}
a_{i\pi-\kappa}\\
b_{i\pi-\kappa} \end{array} \right),(n>0)\\
\end{aligned}
\end{equation}
The evolution equations at $n=\pm1$ become the equations of $\kappa$ and $t$, a system of non-linear equations,
\begin{equation}
\begin{aligned}
\omega({\cos\theta}ta_{i\pi-\kappa}+{\sin\theta}b_{\kappa})&={\cos\theta}a_{\kappa}+{\sin\theta}b_{\kappa}\\
\omega({\sin\theta}ta_{i\pi-\kappa}-{\cos\theta}b_k)&=t({\sin\theta}a_{i\pi-\kappa}+{\cos\theta}b_{i\pi-\kappa})\label{Eq. 30}
\end{aligned}
\end{equation}
The four solutions in Eq.~\ref{Eq. 14} are obtained from solving these two equations.

\end{document}